# Testing an instrument to measure the BPMS-KM Support Model

Alicia Martín-Navarro
INDESS, Universidad de Cádiz, Spain
alicia.martin@uca.es

María Paula Lechuga Sancho*
INDESS, Universidad de Cádiz, Spain
paula.lechuga@uca.es
*Corresponding Author

José Aurelio Medina-Garrido
INDESS, Universidad de Cádiz, Spain
joseaurelio.medina@uca.es

## ABSTRACT

BPMS *(Business Process Management System*) represents a type of software that automates the organizational processes looking for efficiency. Since the knowledge of organizations lies in their processes, it seems probable that a BPMS can be used to manage the knowledge applied in these processes. Through the BPMS-KM Support Model, this study aims to determine the reliability and validity of *a 65-item instrument* to measure the utility and the use of a BPMS for knowledge management (KM). A questionnaire was sent to 242 BPMS users and to determine its validity, a factorial analysis was conducted. The results showed that the measuring instrument is trustworthy and valid. It represents implications for research, since it provides an instrument validated for research on the success of a BPMS for KM. There would also be practical implications, since managers can evaluate the use of BPMS, in addition to automating processes to manage knowledge.

## INTRODUCTION

Since the 1980s, Business Process Management (BPM) has been a topic intensely discussed, in research on information systems (Houy et al., 2010). In previous decades, many organizations sought technological initiatives that would allow them to make changes, manage their businesses and improve their performance (Harmon, 2010). These initiatives had extended across departmental and organizational boundaries, including customers and suppliers, resulting in a transformation of functional organizations to organizations by processes. The growing interest in improving the management of processes in the organization entails the development of different types of information systems to support to the management of processes. Most innovative of the systems developed for this purpose is the BPMS, oriented to flexible management and continuous optimization of the processes (Wong, 2013). The BPMS represents a type of software that allows one to manage the business processes of an organization (Aalst et al., 2003), through its design and modeling (Smith & Fingar, 2003). Therefore, BPMS allows an organization to more quickly adapt to the continuous changes of the market and its consumers. The companies that adopt this technology are interested in the value that these systems generate through the continuous improvement of the business processes, improving their competitive position (Wong, 2013).

On the other hand, the search for competitive advantages through the continuous improvement of processes has awakened the interest of academics and executives to KM (Wang & Yang, 2016). KM is a term that has existed since the 1980s and can be defined as a systematic and deliberate creation, an update and a use of knowledge to maximize the efficiency of the organization (Slavacek, 2011). Among others, questions related to individual knowledge, outsourcing, formalization (for example, incorporating it within a process) and diffusion in organizations (Kalpic & Bernus, 2006) have been heavily researched.

BPM and KM have been widely developed as independent disciplines. Nevertheless, for an organization that includes both issues in the improvement of its processes, the two should be managed in a coordinated way (Slavacek, 2011). Indeed, BPMS offers the opportunity to improve BPM and KM simultaneously. However, the literature does not test variables that allow one to evaluate the success in the use of a BPMS for such a goal. Although there are diverse works that define custom scales to measure the success of other information systems, such as *Enterprise Resource Planning* (ERP) (Stratman and Roth, 2002; Hou, 2014) or *Knowledge Management System* (KMS) (Kulkarni, Ravindran, & Freeze, 2006), so far no scale has been found that measures the success in the use of the BPMS for KM. This gap in the literature highlights the need to develop a latent construct capable of measuring this success as close as possible to reality. The measurement of the success of a BPMS in the management of processes and knowledge requires the development of a theoretical model that, in this work, we will call *the BPMS-KM Support Model*, and of the measuring instruments of the variables to include in this model.

The objective of this work is to define and validate a scale of measurement of the variables that, according to literature, can influence the success of the use of a BPMS for KM, and that will form part of the so-called *BPMS-KM Support Model*, determining the reliability and validity of the proposed measuring instrument.

For the attainment of this objective, this work is structured in the following way. After this introduction, a review of the most relevant theoretical framework for this study is conducted. Section 3 specifies the methodology used, as well as the design of the measuring instrument and the data collection process. Section 4 discusses the results of the validation of the proposed scale. Finally, Section 5 presents the main conclusions drawn and exposes the limitations of the work, as well as the future research agenda.

## BACKGROUND

Over the past few decades, many organizations implemented information technologies to manage their business processes with the intention to improve their performance (Harmon, 2010). These technologies extended the functional and organizational limits, including customers and suppliers, transforming the organization by functions into an organization by processes. Against this backdrop the BPMS arose (Wong, 2013), to overcome the limitations of the existing technology, which was not flexible and could not be easily adapted to organizational changes and processes. The BPMS is a graphical and flexible design of models that represent the processes, showing visually the sequential flow of related activities within each process. In addition, it allows agile management of these processes, adapting quickly to them, monitoring them, improving them and optimizing them continuously (Smith, Fingar, & Scott, 2002). Thanks to the BPMS, changes in the processes become quickly operational, and the work can be automated without needing programming. These systems are not meant to replace existing applications within the organization; they add the use of these applications to a new process and use the information, being able to integrate all this information in a manner that is more flexible than if they were to reprogram the software (as if it happens with *workflows*), thus the businesses becomes more agile (Wong, 2013). This facilitates the integration of the BPMS with other heterogeneous systems that are constituted as a platform for the development of applications based on processes (Rhee, Bae, & Choi, 2007; Smith & Fingar, 2003).

On the other hand, the organizational knowledge is a valuable resource that should be incorporated in the processes (Rastogi, 2002). The knowledge that resides only in the mind of a person has limited value, whereas that value can be increased exponentially when it is redirected, reused and integrated into the processes (Douglas, 2002). Knowledge is a strategic resource that should be constantly be identified, measured, acquired, developed, used and protected (Bitkowska, 2015). The competitiveness of the organizations will depend on how they apply, exploit and integrate knowledge (Alavi & Leidner, 2001; Šajeva, 2010). The need to manage it properly increases the importance of the KM concept. A suitable KM can help improve practices of work, the efficiency of the processes, the decision making and service to the customer; allowing to shorten development time, train employees, innovate, increase sales and create value (Dotsika & Patrick, 2013; Šajeva, 2010). Therefore, the organizations must manage and systematically deliberate their knowledge base (Han & Park, 2009).

Wu, Kao & Chen (2015) indicate that knowledge creation is strongly linked, among other aspects, to the processes. If the knowledge of an organization is used in the processes, the BPMS that automate these processes could play an important role in KM. In this sense, BPMS and KM are two interconnected paradigms that can generate synergies when, for example, the use of certain knowledge in a process is standardized when incorporating it in a BPMS. Thus, Kalpic & Bernus (2006) defend that a BPMS, designed specifically for the efficient management of processes, is also an important tool for KM, which allows the transformation of informal knowledge into formal knowledge and facilitates its externalization, distribution and subsequent internalization. Those users of a BPMS who generate knowledge in their daily tasks will probably incorporate the knowledge generated to the processes and their own BPMS.

## BPMS-KM SUPPORT MODEL

The literature recognizes the interest to evaluate the success of the BPMS (Alam et al., 2015; Antonucci & Goeke, 2011; Bai & Sarkis, 2013; Bălănescu et al., 2014; Filipowska et al., 2009; Gabryelczyk, 2016; Janssen et al., 2015), and as it occurs with other information systems, many authors have used the "system use" and the "user satisfaction" as measures of success. More specifically, one of the most widely used models in the academic literature, to measure the success of an information system, is the ISSM (*Information System Succes Model)* (Asmah, Ofoeda, & Gyapong, 2016; Chang, Chang, Wu, & Huang, 2015; DeLone & McLean, 1992; Jiang & Wu, 2016; Poelmans et al., 2013; Sultono, Seminar, & Erizal, 2015; Urbach, Smolnik, & Riempp, 2011; Wu & Wang, 2006; Yusof & Yusuff, 2013). In 1992, DeLone & McLean conducted a literature review in which they identified six different aspects or categories of information systems from which they made up the ISSM model, these are: (1) system quality, (2) information quality, (3) system use, (4) user satisfaction, (5) individual impact and, (6) organizational impact. Ten years after the publication of this first model and based on the assessment and contribution of several researchers (Kettinger & Lee, 1994; Pitt, Watson, & Kavan, 1995; Seddon, 1997), Delone & McLean (2003) re-specified the original model resulting in the Update IS Succes Model (see figure 1). This new model is focused on end users and many researchers have used it to assess different information systems, including technologies that were not widely used when the model was created, such as electronic documentation systems (AlShibly, 2014), cloud computing (Lian, 2017), knowledge management systems (Wu & Wang, 2006), *e-learning* systems (Lin, 2007b), *e-government* systems (Wang, 2006), and even BPMS (Poelmans et al., 2013).

Figure 1. Update ISSM

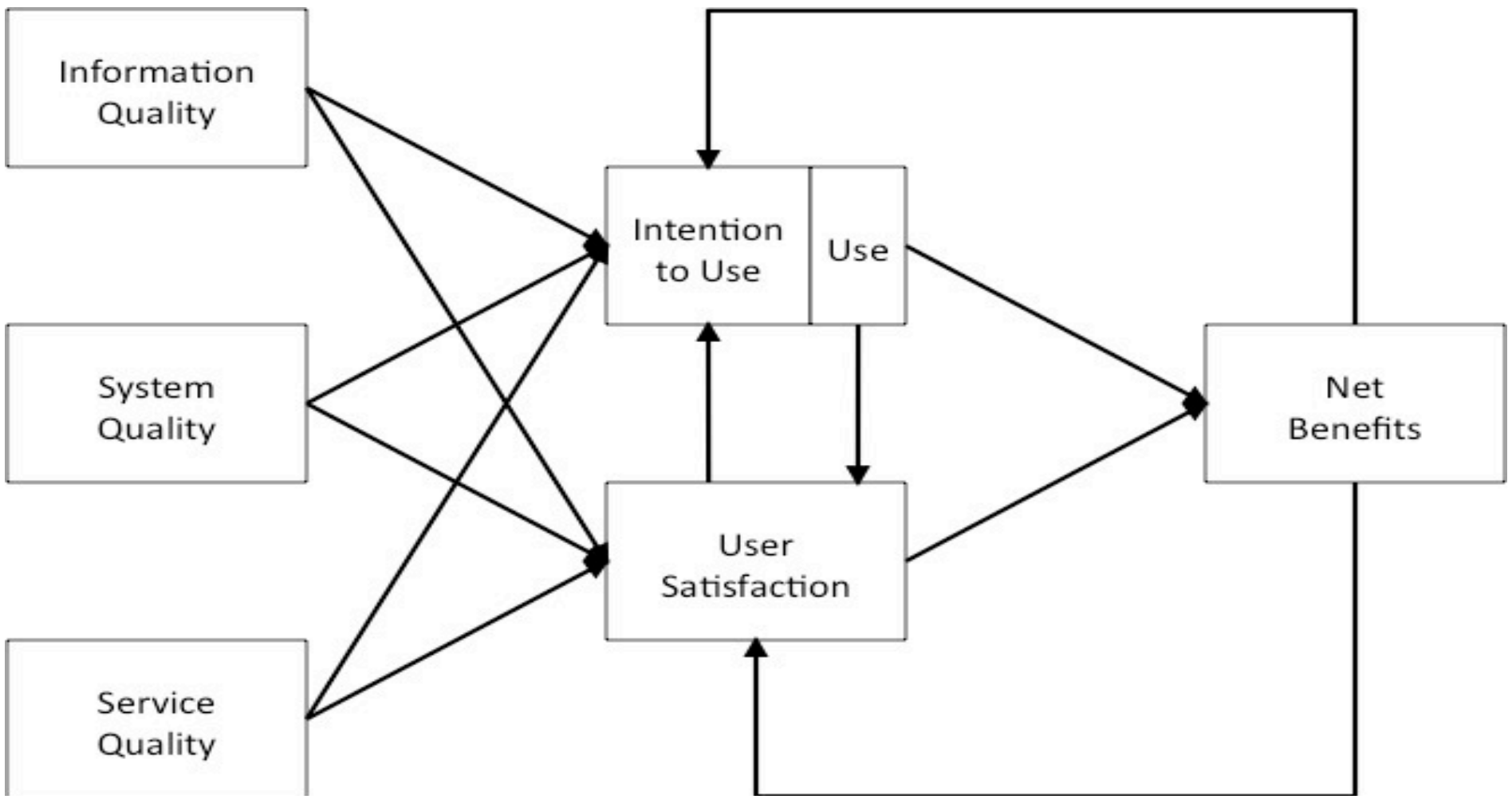

Source: Delone & McLean (2003)

From the Update ISSM, Wu & Wang (2006) created the success model of a knowledge management system (KMS). As is the case of most information systmes, the success of a KMS depends, in part, on the degree of use (Poston & Speier, 2005) and, in turn, it must be tied to the quality of the system, the quality of the information, user satisfaction and perceived usefulness. In this way, technological dimensions such as system and information quality, and human dimensions such as user satisfaction, perceived benefits and use of the system can be considered as appropriate constructs to measure the success of a knowledge management system (Wu & Wang, 2006). This research led to what is now known as the KMS success model. Using the Update ISSM as theoretical background, they introduced information and knowledge quality as a measure of success. In addition, they also developed new measurement scales for the previous variable and for the system use, in a context of knowledge management systems. Likewise, the success model for BPMS, developed by Poelmans et al. (2013), was inspired by the Updated ISSM (Delone & McLean, 2003). This other model measures the operational success of these systems through perceived utility and user satisfaction with the tool. Poelmans et al. (2013), found as background of both constructs the System Dependency, System Quality and Information Quality. Regarding the technological part of the software, they identified four dimensions in System Quality, which were considered a multidimensional construct: (1) Service Quality, assessed through the training and support of the IT department, (2) Input Quality, of great importance because an information system is an input-processing-output system and therefore the user not only evaluates the software by the quality of the information it generates, but also by how the information may be loaded, (3) General System Attributes are attributable to any information system, and are measured by Responsiveness, Reliability and Integration; and (4) BPMS-specific System Attributes which measures BPMS unique features, namely, Allocation and Routing. The definition of these constructs can be checked in table I.

As indicated above, processes and knowledge are closely related. In this way, BPMS, as a system that automates processes well, must integrate user knowledge into the software. Therefore, the integration of the variables contemplated in the referred models of success applied to the KMS and to the BPMS allow us to propose a new model, which we will call the *BPMS-KM Support Model* (see Figure 2), in that it analyzes the success of the use of a BPMS for KM.

Figure 2. *BPMS-KM Support Model*

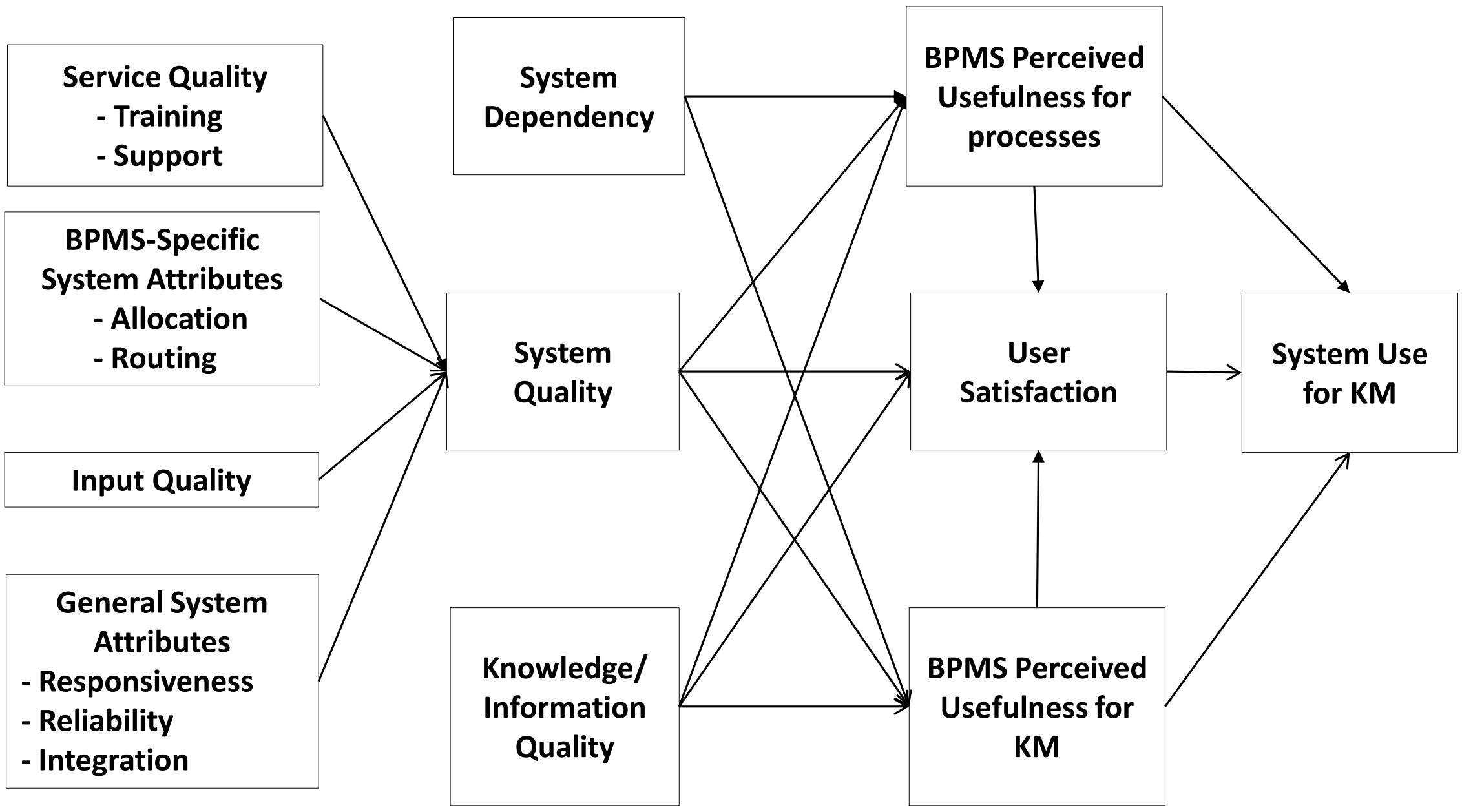

Source: Own elaboration from Wu & Wang (2006) and Poelmans et al. (2013)

*The BPMS-KM Support Model* is therefore developed following Delone and McLean´s call for further development and validation of their model, and it consists on eleven variables whose reliability and validity will be determined in this work through a factorial analysis. These variables are: *service quality; input quality; system quality; knowledge/information quality; BPMS-specific and general system attributes; system dependency; perceived usefulness for processes and for KM; user satisfaction; and system uses for KM.* In Table I, there is a definition of each one of the considered variables.

Table I. Construct definitions

| Construct | Definition |
|---|---|
| Service quality | Quality of support to user receives from system specialist (Ramirez-Strap, P., Alfaro-Perez, J., & Cancino-Flower, 2015) |
| BPMS-specific system attributes | Routing quality and allocation are two BPMS specific features. Routing quality measures the options to send a case forward or backward along a process. Allocation quality assesses the selection of work items that BPMS makes within a particular step in the business process (Aalst & Hee, 2004) |
| Input quality | Degree to which a BPMS allows the user to enter data in a complete, understandable, sufficient, relevant and correct way, and at the right moment (Poelmans et al., 2013). |
| General system attributes | Reliability, responsiveness and integration are relevant to all types of information systems. Reliability refers to the user's dependence on the operative of the system. Responsiveness is the response time offered by the system. And integration makes reference to how the system integrates data from different sources (Wu & Zhang, 2014) |
| System dependency | Degree of interaction the user needs with the system (Poelmans et al, 2013) |
| System quality | Accessibility and flexibility of the system, taking into account the ease of use and availability of the same (Wixom & Todd, 2005). |
| Knowledge/information quality | Degree to which the system generates information in a sufficient and appropriate way (Delone & McLean, 2003). |
| Perceived usefulness for processes | Perception that a person having to use the system will improve the task performed (Davis, 1989). |
| User satisfaction | The sum of feelings of pleasure or displeasure toward the benefits that a user expects to receive from the interaction with an information system (Seddon, 1997). |
| Perceived usefulness for KM | Perception that a person having to use the system will improve the information and knowledge (Davis, 1989). |
| System use for KM | Degree to which the system is being used for KM (Wu & Wang, 2006). |

Because when evaluating the success of a BPMS, the opinion of an end user is consolidated as an accepted measurement, it becomes necessary the analysis of the factors that determine the use of the BPMS for KM will indicate the viability of this system for KM. Although there are no validated models in the literature to analyze the success of the use of BPMS for KM, there are proven success models for the Knowledge Management Systems (KMS) (Wu & Wang, 2006) and for the BPMS (Poelmans et al., 2013) separately, and none of the models would be valid for studying the use of the BPMS for KM. Therefore, it seems unavoidable to combine both models for researchers to evaluate the capability of a BPMS for KM, through the *BPMS-KM Support Model.*

## METHODOLOGY

The validation of the measurement of the variables of any theoretical model is a relevant and inherent aspect in all scientific research (Coronado Padilla, 2007; Mendoza & Garza, 2009). In this context, the design and validation of scales is intended to develop formal models to make viable the measurement of abstract and non-observable concepts (Martinez-Arias, 2005), which can only be measured indirectly, through indicators (Agudo et al., 2010).
According to Churchil (1979) and De Vellis (1991), the process of design and validation of scales is grouped in seven stages listed in Figure 1. These stages will be developed in this methodological section to validate the scale that measures the variables of the construct represented by the *BPMS-KM Support Model* proposed in the previous section. This model, in fact, covers stage one: "Definition of construct."

Figure 3. Process of design and validation of a measurement scale

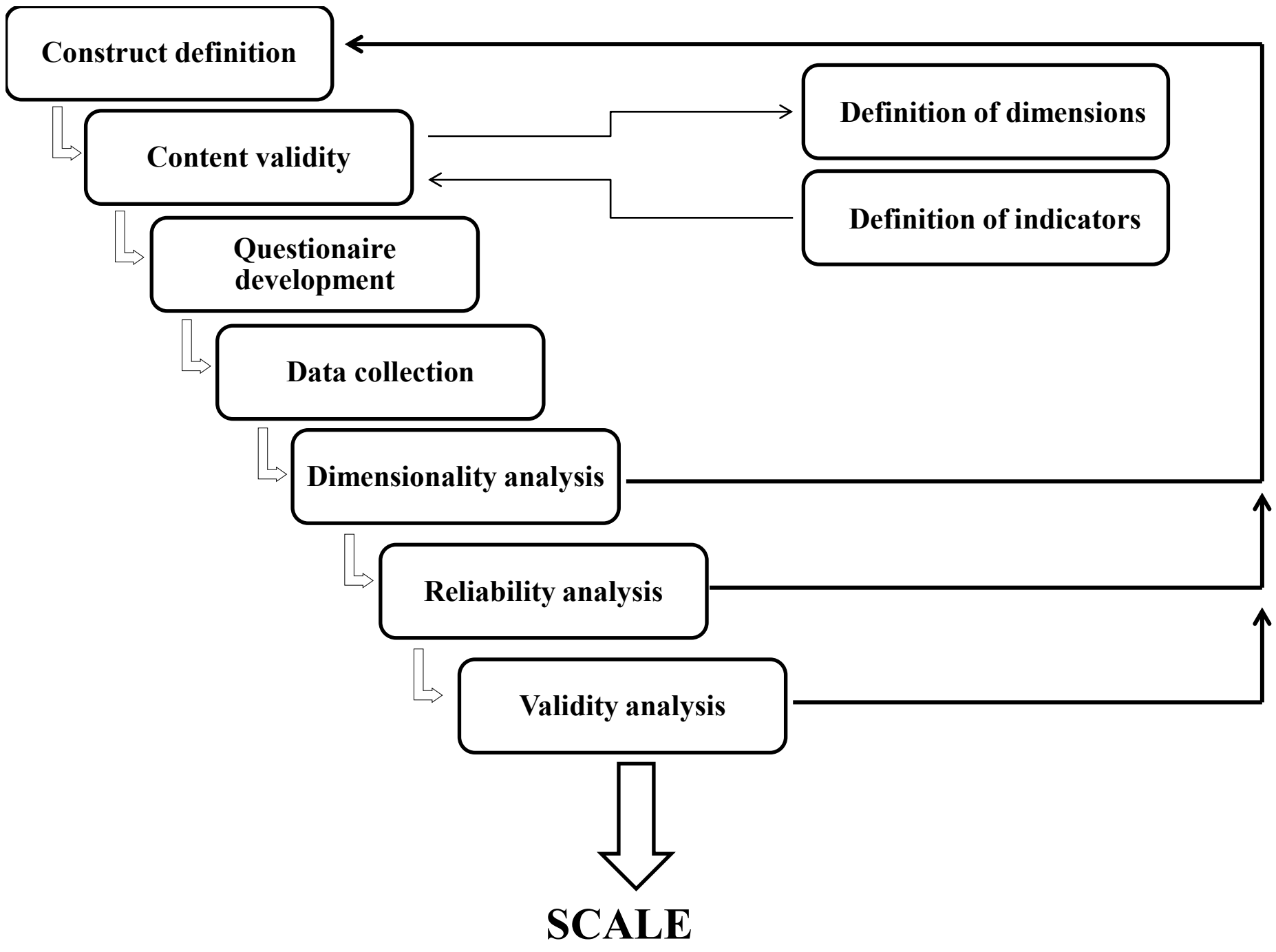

Source: Adapted from Churchill (1979), De Vellis (1991), Hair et al., (2014).

**Validity of content**

According to Bigné (1999), content validity studies if the scale picks up different aspects or dimensions that are considered basic and fundamental in relation to the object of analysis, that is to say, construct or latent variable object of study (Mulero, 2010). In this sense, to determine if an instrument has validity of content, it is necessary to verify if the elements that constitute the same define well the conduct to measure (Haynes et al., 1995). Therefore, the items must be relevant and representative of the dimensions or contents of the construct object of study for a specific purpose of measurement (Haynes et al., 1995; Netemeyer et al., 1996). This type of validity cannot be measured numerically, but subjectively by researchers (Bollen, 1989). Similarly, Hair et al. (2014) affirmed that content validation serves to subjectively appreciate the correspondence between the individual items and the concept.

At this stage, the researcher must consider if the questionnaire to use has been designed specifically for the research, if it is a modified questionnaire of other authors, or if the questionnaire was completely designed by another researcher (Mendoza & Garza, 2009). Specifically, this study uses a questionnaire that has resulted from modification of others designed by different authors. These modified questionnaires are mainly those that were designed for the studies of Poelmans et al. (2013) and Wu & Wang (2006) and, therefore, the variables have been previously justified in the literature to define them within the scope of the BPMS (see Table I). However, the new instrument proposed in this work does not fully reflect the validity and consistency of the original works, so it becomes fundamental to re-establish its reliability and validity (Mendoza & Garza, 2009).

**Elaboration of the questionnaire**

As indicated above, the questionnaire designed in this research is constructed from other existing questionnaires in the literature on the paradigm of information systems and, particularly, the BPMS (Poelmans et al., the 2013) and the KM (Wu & Wang, 2006). Given the defined theoretical construct, the questionnaire was designed for the end users of a BPMS.

Once the initial design was completed, the questionnaire was reviewed by a panel of experts formed by researchers and professionals of the sector, with the aim of guaranteeing its formal quality and of content. Among the professional experts of the sector were three members of the formation team of a software development company, specializing in BPMS. Also included on the panel were two modelers that created work paths through a BPMS and who worked for companies that have this implemented software. The contact with the professionals was carried out through informal interviews. As a result of the interviews, the questionnaire was improved to make it more comprehensible for the end users. Once the debugging process was concluded, and following Lee, Tsao, & Chang (2015), the final survey was structured into two large distinct blocks. The exact transcription of all the items incorporated into the questionnaire can be found in the Appendix.

The first block collected demographic variables and was composed of ten items about the user, specifically gender, age, country of origin, academic level, economic sector to which their company belongs, and number of employees. Next, information was successfully obtained on the interaction of the user with the system. To do this, they were asked to indicate the experience that they had with the system after months of use. In addition, those surveyed indicated the process for which they used the BPMS most.

The second block of the questionnaire gathered eleven sections amounting to a total of sixty-five items, measured on a seven-point Likert scale (Dittrich, Francis, Hatzinger, & Katzenbeisser, 2005; Kwok, 2014). Each one of the sections included a brief explanation so that the users could better understand the questions. The items considered the hours weekly that the users invested in the system ("System Dependency" measured by item variable); the training and support received for the proper use of the system ("Service Quality", with four items); a subjective evaluation of the general and specific attributes of the system used ("BPMS-Specific Attributes System", with four items; and "General System Attributes", with seven items); the options that the system showed them to enter information ("Input Quality", with six items); the quality of the system in operative terms ("Quality System", with five items); and the quality of the information and knowledge that the system provided them ("Knowledge/Information Quality", measurement with twenty items). Respondents were also asked to evaluate the benefits they perceived they gained from the software to perform the processes ("Perceived Usefulness for Processes", with four items) and KM ("Perceived Usefulness for KM", with five items). Other questions made reference to the feelings of affability or displeasure with the system ("User Satisfaction", with four items). Finally, they were asked about using the system for KM ("System Uses for KM", with five items).

**Data collection**

Data collection was conducted through a directed study. This was necessary because the studied system is implemented only in a small number of companies. This type of study is common in quantitative empirical research in the area of the information systems (Chang, Chang, Wu, & Huang, 2015; Feng, Tien, Feng, & Lai, 2014; Hariguna, Lai, & Chen, 2016; Kulkarni, Ravindran, & Freeze, 2006; Mudzana & Maharaj, 2015; Poelmans et al., 2013; Urbach, Smolnik, & Riempp, 2011; Wu & Zhang, 2014; Wu & Wang, 2006).

To perform data collection, we designed an electronic form of Google that allowed the respondent to autoadministrate the questionnaire online, in accordance with what has been done in similar academic works (Elbashir, Collier, & Davern, 2008; Feng, Tien, Feng, & Laid, 2014; Hariguna, Lai, & Chen, 2016; Jiang & Wu, 2016; McGill & Klobas, 2009; Mudzana & Maharaj, 2015; Poelmans et al., 2013; Wu & Zhang, 2014). This type of method is very useful since it allows the users to access the survey through a web link. In addition, the collected data goes directly to a spreadsheet, which facilitates the subsequent statistical treatment of the data. The collection of data was conducted during the months of November and December of 2016, and January and February of 2017.

Unable to find a record of companies that have implemented BPMS, we proceeded to identify such organizations through the websites of suppliers. With all this information, a database was generated with a total of 53 companies that we attempted to contact to explain the research. We obtained the participation of 12 Spanish and Latin American companies, which were sent a total of 415 questionnaires to disseminate amongst their employees who use BPMS. To make the dissemination process easier, a standard letter was created directed to these end users. The letter briefly explained the project and its objective, and provided them with the link to the form. End users of different BPMS completed a total of 242 surveys. These BPMS were mainly AuraPortal BPM, Cibernos Plan, Bizagi and SAP.

The descriptive data of the study is shown in Table II. As observed in this table, the respondents have extensive experience using the BPMS implemented in their companies (35.57 months, almost 3 years on the average) and certain degree of dependency on these systems, measured by the number of hours per weekly using them (10.29 hours of average). Of the 242 surveyed individuals, 63.6% are men and 36.4% are women. The sample has been divided into four categories by age as follows: under 25 years old (2.5% of respondents), 25 to 40 years old (55.8%), 41 to 55 years old (36.4%) and over 55 years old (5.4%). By educational level, it is observed that the 0.4% of the respondents only have primary education, 8.1% have secondary or vocational training, 75.2% have a college degree and the 15.7% have postgraduate degrees. It is interesting to note that most survey users of BPMS, more than 90%, have higher education. With respect to the activity sector, 16.1% of the respondents work in the productive sector, 25.9% in the financial sector, 26.9% in the services sector and 31.4% in other sectors. On the other hand, in relation to the size of the participant companies, considering the number of employees, 13.2% of the respondents work for companies with less than 50 employees, 15.3% with between 51 and 100 employees, 25.6% with between 101 and 500 employees, 23.1% with between 501 and 1000 employees, and 22.7% with more than 1000 employees.

Table II. Organizational and demographic descriptive data of the respondents.

| | | N | % | Average |
|---|---|---|---|---|
| BPMS experience (in months) | | | | 35.57 |
| BPMS dependency (hours per week) | | | | 10.29 |
| Gender | Male | 154 | 63.6 | |
| | Female | 88 | 36.4 | |
| Age | Under 25 | 6 | 2.5 | |
| | 25-40 | 135 | 55.8 | |
| | 41-55 | 88 | 36.4 | |
| | Over 55 | 13 | 5.4 | |
| Education level | Primary | 1 | 0.4 | |
| | Secondary | 21 | 8.1 | |
| | Grade | 182 | 75.2 | |
| | Postgraduate | 38 | 15.7 | |
| Industry | Manufacturing | 39 | 16.1 | |
| | Financial sector | 62 | 25.9 | |
| | Services | 65 | 26.9 | |
| | Others | 76 | 31.4 | |
| Company size (number of employees) | 1-50 | 32 | 13.2 | |
| | 51-100 | 37 | 15.3 | |
| | 101-500 | 62 | 25.6 | |
| | 501-1000 | 56 | 23.1 | |
| | Over 1000 | 55 | 22.7 | |

To determine the normal distribution of the data in this study, the sample was examined through the values of asymmetry and kurtosis. All the values of the items are les than two for asymmetry and less than seven for the kurtosis (Curran, West, & Finch, 1996). In view of these statistical results, it is understood that there is univariate normality of the data of the study.

**Analysis of dimensionality**

Understanding the scale as a set of items, the unidimensionality of it refers to only a single latent characteristic or non-observable construct located at the base of it (Hattie, 1985). Therefore, the purpose of studying the unidimensionality is to verify if the scale, as a whole, and the possible subscales, have their own organization by themselves and if, with their scores, they represent a single dominant factor.
One of the statistical techniques used most to evaluate the unidimensionality of a construct is the factorial analysis (Muñiz Fernandez, 1997; Perez-Gil, Chacón Moscoso, & Moreno Rodríguez, 2000). The application of any factorial analysis requires the previous study of the homogeneity of the set of items included on the scale. For this reason, two tests are usually used: the sample adequacy test of Kaiser Meyer Olkin (statistical KMO) and the test of sphericity of Bartlett. The KMO indicates how broad the correlation is between the measurement variables and varies between 0 and 1. From 0.5, the adequacy of the sample for a factorial analysis is deemed good. On the other hand, Bartlett's test of sphericity verifies that the matrix of correlations adjusts to the identity matrix and, therefore, there is no significant correlation between the variables. Its value must be below 0.01. Table III refers to the values of the KMO test, Bartlett's test of sphericity and the percentage of variance explained. As it is observed, the statistical KMO is sufficiently high in all the cases, in accordance with the recommended. In addition, the significance of Bartlett's test of sphericity is, in all the cases, equal to zero. It is observed, therefore, perfect adequacy of the data to a model of factorial analysis.
In terms of the percentage of explained variance, a listing of the eigenvalues of the variances-covariances matrix and the percentage of variance is offered that represents each of them. The eigenvalues express the amount of total variance that is explained by each factor. By default, many factors, such as eigenvalues greater than 1, are analyzed and extracted. In order to verify the unidimensionality, the criterion of Kaiser (1960) is applied; an eigenvalue must exist and be greater than 1. Following Hair et al., (2014), with respect to the percentage of explained variance, an approximated percentage of 60% can be considered satisfactory in the field of social sciences. This explained variance allows us to affirm that there exists interrelation between the items of each factor in each one of the constructs.

Table III. Unidimensionality of the scale.

| | | KMO | Barlett's sphericity test | | | % Variance explained |
|---|---|---|---|---|---|---|
| Construct | Labels | | Chi-Square | df | Significance | |
| Service quality | SERVIQ | 0.806 | 1014.613 | 6 | 0.000 | 86.101 |
| General system attributes: reliability | RELIAB | 0.692 | 234.354 | 3 | 0.000 | 71.457 |
| General system attributes: responsiveness | RESPON | 0.5 | 473.065 | 1 | 0.000 | 96.402 |
| General system attributes: integration | INTEGRA | 0.5 | 312.69 | 1 | 0.000 | 92.690 |
| BPMS-specific system attributes | SPECIFIC | 0.794 | 505.649 | 6 | 0.000 | 71.296 |
| Input quality | INPUTQ | 0.923 | 1318.755 | 15 | 0.000 | 78.707 |
| System quality | SYSTEMQ | 0.844 | 1047.303 | 10 | 0.000 | 77.140 |
| Information and knowledge quality: retrieval quality | IKQRETRI | 0.922 | 2322.317 | 36 | 0.000 | 74.633 |
| Information and knowledge quality: content quality | IKQCNTN | 0.915 | 1999.949 | 21 | 0.000 | 81.953 |
| Information and knowledge quality: context and linkage quality | IKQCNTX | 0.702 | 746.334 | 6 | 0.000 | 73.826 |
| Perceived usefulness for processes | USEFULP | 0.842 | 1029.312 | 6 | 0.000 | 87.593 |
| Perceived usefulness for KM | USEFULK | 0.788 | 1022.164 | 6 | 0.000 | 85.602 |
| User satisfaction | SATISF | 0.922 | 1996.16 | 10 | 0.000 | 92.824 |
| System use for KM | USE | 0.852 | 1426.26 | 10 | 0.000 | 85.090 |

**Analysis of the reliability**

To examine the adequacy and the validity of the construct of the proposed measurement model, we evaluated the measuring instrument through the reliability of the scale, the convergent validity and the discriminating validity. As far as the reliability, this was evaluated using composite reliability and the alpha of Cronbach ($\alpha$), which allowed for the assessment of the internal consistency of all the indicators that measure the constructs. Table IV shows the collection of these statistics for all the constructs. Cronbach's alpha coefficient has its ideal levels, between 0.8 and 0.9 (Streiner, 2003). As observed in the table, all of the constructs widely surpass the threshold required for Cronbach's alpha, except, the construct of general attributes of the system (reliability). Nevertheless, it has a value that is close to 0.8, only missing it by three thousandth. Therefore, it could be considered acceptable and admitted being a value so close (Cronbach & Meehl, 1955). The index of *composite reliability* (Fornell & Larcker, 1981) is interpreted as the alpha of Cronbach, but it considers the interrelations of the extracted constructs. This statistical value is recommended to be greater than 0.7 to be an acceptable level. However, if one wants high levels, the composite reliability would have to be around 0.9 (Prieto & Delgado, 2010). It is observed that the *composite reliability* has some very high values for each one of the constructs.

Table IV. Reliability indicators

| Construct | Labels | Cronbach's alpha ($\alpha$) | Composite reliability (CR) |
|---|---|---|---|
| Service quality | SERVIQ | 0.945 | 0.961 |
| General system attributes: reliability | RELIAB | 0.797 | 0.882 |
| General system attributes: responsiveness | RESPON | 0.936 | 0.982 |
| General system attributes: integration | INTEGRA | 0.920 | 0.962 |
| BPMS-specific system attributes | SPECIFIC | 0.854 | 0.908 |
| Input quality | INPUTQ | 0.944 | 0.957 |
| System quality | SYSTEMQ | 0.923 | 0.944 |
| Information and knowledge quality: retrieval quality | IKQRETRI | 0.956 | 0.964 |
| Information and knowledge quality: content quality | IKQCNTN | 0.962 | 0.969 |
| Information and knowledge quality: context and linkage quality | IKQCNTX | 0.880 | 0.918 |
| Perceived usefulness for processes | USEFULP | 0.952 | 0.966 |
| Perceived usefulness for KM | USEFULK | 0.943 | 0.960 |
| User satisfaction | SATISF | 0.980 | 0.985 |
| System use for KM | USE | 0.956 | 0.966 |

**Analysis of the validity**

*Convergent validity*

Validity is examined by observing the convergent validity and discriminant validity of a construct and it is the extent to which the items on a scale measure the abstract or theoretical construct (Carmines and Zeller, 1994; Hair et al., 2014). The existing correlations between the indicators of the same construct are measured through the convergent validity. If the measures are significant, this means that the scale is congruent (Luján-Tangarife & Cardona-Arias, 2015). Following Fornell & Larcker (1981), to determine the convergent validity, the *Average Variance Extracted* (AVE) has been calculated. This indicator measures if the variance of the construct can be explained through the chosen indicators. The recommended values of AVE should not be less than 0.5, according to Bagozzi & Yi (1988). This indicates that more than 50% of the variance of the construct is due to its indicators. Therefore, as shown in Table V, all the constructs have a coefficient greater than 0.5 AVE. Therefore, all the indicators that comprise the same construct are highly correlated to each other and are correctly measuring the study phenomenon.

Table V. Convergent validity

| Construct | Labels | AVE |
|---|---|---|
| Service quality | SERVIQ | 0.8609 |
| General system attributes: reliability | RELIAB | 0.7146 |
| General system attributes: responsiveness | RESPON | 0.9643 |
| General system attributes: integration | INTEGRA | 0.9274 |
| BPMS-specific system attributes | SPECIFIC | 0.7132 |
| Input quality | INPUTQ | 0.7870 |
| System quality | SYSTEMQ | 0.7716 |
| Information and knowledge quality: retrieval quality | IKQRETRI | 0.7464 |
| Information and knowledge quality: content quality | IKQCNTN | 0.8194 |
| Information and knowledge quality: context and linkage quality | IKQCNTX | 0.7385 |
| Perceived usefulness for processes | USEFULP | 0.8757 |
| Perceived usefulness for KM | USEFULK | 0.8559 |
| User satisfaction | SATISF | 0.9279 |
| System use for KM | USE | 0.8507 |

*Discriminant validity*

Discriminant validity exists when there is a low correlation between the scale used to measure a construct and the scales of measurement used for the rest of the considered constructs (Churchill, 1979; Sanchez and Sarabia, 1999; Vila et al., 2000). The criterion to verify the discriminating validity is that the square root of the AVE of the construct is greater than the correlation between that construct and all the others (Chin, 1998). Table VI shows the coefficients of correlation between the constructs (see labels of the items in the Appendix). In the diagonals, instead of the classic value of 1, the square root of the AVE is shown. Specifically, it is observed for all the cases that the existing correlations between the items within the same construct are higher than the existing correlations between different constructs. From the above, it is possible to admit that the questionnaire has discriminant validity and, therefore, no item measures a construct different from which it belongs.

Table VI. Discriminant validity.

| | SERVIQ | RELIAB | RESPON | INTEGRA | SPECIFIC | INPUTQ | SYSTEMQ | IKQRETRI | IKQCNTN | IKQCNTX | USEFULP | USEFULK | SATISF | USE |
|---|---|---|---|---|---|---|---|---|---|---|---|---|---|---|
| SERVIQ | .9278 | | | | | | | | | | | | | |
| RELIAB | .440 | .8453 | | | | | | | | | | | | |
| RESPON | .540 | .728 | .9820 | | | | | | | | | | | |
| INTEGRA | .520 | .486 | .540 | .9630 | | | | | | | | | | |
| SPECIFIC | .646 | .569 | .601 | .729 | .8445 | | | | | | | | | |
| INPUTQ | .614 | .593 | .652 | .676 | .817 | .8879 | | | | | | | | |
| SYSTEMQ | .588 | .588 | .609 | .594 | .699 | .772 | .8784 | | | | | | | |
| IKQRETRI | .590 | .600 | .663 | .627 | .758 | .834 | .800 | .8639 | | | | | | |
| IKQCNTN | .565 | .544 | .618 | .669 | .758 | .813 | .814 | .859 | .9052 | | | | | |
| IKQCNTX | .498 | .443 | .480 | .588 | .618 | .661 | .651 | .630 | .738 | .8594 | | | | |
| USEFULP | .588 | .510 | .570 | .608 | .743 | .767 | .783 | .854 | .842 | .694 | .9358 | | | |
| USEFULK | .605 | .446 | .491 | .554 | .672 | .668 | .675 | .723 | .769 | .750 | .812 | .9252 | | |
| SATISF | .580 | .524 | .560 | .603 | .702 | .769 | .799 | .808 | .792 | .716 | .869 | .856 | .9633 | |
| USE | .534 | .345 | .363 | .499 | .583 | .563 | .546 | .597 | .656 | .655 | .687 | .770 | .702 | .9223 |

## RESULTS

This study has combined two different models into a final model called BPMS-KM Support Model. The first of these is the successful model of a knowledge management system (KMS) (Wu and Wang,2006). This model is used to evaluate the success of a technology for knowledge management, developing specific constructs and scales for that purpose. The second model evaluates the operational success of a BPMS (Poelmans et al., 2013) and includes specific process automation technologies constructs. Because neither of them, separately, can be used to measure the success of a KM BPMS, the two have been combined. This combination and integration generates the so called BPMS-KM Suppot Model, and a questionnaire has been developed with specific constructs and scales of a BPMS in a KM context.

The validity of content of the questionnaire is justified by the integration of questions pertaining to other questionnaires validated by previous literature. Thanks to the results of the asymmetry and kurtosis, it was possible to determine that the items follow a normal distribution. Also, through factorial analysis, KMO indicators and the sphericity of Bartlett, it was shown that there really is unidimensionality of the scales. The variance explained shows that there is interrelation between each factor and items, through the subsequent matrix of main components, leaving one to conclude that each one of the items exactly measured the construct to which it belongs and does not measure any other. The next step consisted of verifying the reliability and the validity of the measuring instrument. One of the indicators used most to measure the reliability is the alpha of Cronbach, which is then reinforced with the composite reliability index. Through these indicators, it has been verified that the scales the questionnaire used are trustworthy. Therefore, it possible to verify good internal consistency of all the indicators that measure the constructs. Finally, convergent and discriminant validity was verified. The convergent validity is calculated through the AVE indicator of each construct and verifies that the AVE values are greater than 0.5 (Bagozzi & Yi, 1988), meaning that the indicators of the same construct are highly correlated to each other. On the other hand, to calculate the discriminant validity, the square root of AVE was used (Chin, 1998) to determine that the existing correlation between the indicators of the same construct are higher than with different constructs. In short, eleven variables have been analysed through a 64-item questionnaire (see appendix) that have successfully undergone its reliability and validity process.

## CONCLUSIONS

The greatest contribution of this work is the development of a series of constructs that measure the success of the use of the BPMS for KM, with a rigorous and validated measuring instrument. For it, we conducted a survey with a sample of 242 end users of BPMS from commercial and private companies. The viability of the questionnaire used has been tested through factorial analysis, verifying the validity and reliability of the measurement scale.

Therefore, we determined that the questionnaire has convergent and discriminant validity. The results allow us to conclude that this questionnaire is a really valid and reliable tool to determine, from the perspective of an end user, whether a BPMS is a useful tool for managing knowledge in organizations.

*Implications*
The results obtained have many implications for theory and practice. This scale has potential value for researchers in BPMS and its application to KM. The items validated for the measurement of the variables of the BPMS-KM Support Model will contribute to research on the determinants of the use of a BPMS to generate, store, formalize or distribute knowledge throughout the organization. This work provides a measurement tool that serves to cover a gap that exists in the academic literature, as no tool or scale has been found to measure the success of how knowledge embedded in processes is managed when these processes are automated with a BPMS. The results obtained cover this gap and imply an advance in relation to other previous analyses developed in the academic literature. In this regard, the validated questionnaire will serve in future research to evaluate the success of a BPMS for KM. From a practical perspective, this study will provide direction for organizations to address whether a BPMS can be used successfully not only to automate their processes, but also to manage knowledge.

*Limitations and future agenda*
The study has some limitations. The sample used in this study comes from commercial companies in Spain and Latin America. For this reason, cultural and economic differences can exist and as far as the implemented BPMS, in relation to other territories. Therefore, the obtained results require further comparisons with other countries and different BPMS. Also, the lack of a robust theory about the use of the BPMS for KM was a limitation for this study. This forced us to use items for the questionnaire validated by studies with other different objectives and technologies. Given these limitations, we propose future lines of research that surpass them. Therefore, it would be interesting to apply this study on the BPMS for KM to other samples originating from different contexts, cultures or countries, as well as samples from the scope of public administrations. Finally, the analysis presented, which has validated the items that measure the variables, represents a prior step to the analysis of *the BPMS-KM Support Model* with models of structural equations that demonstrate the relations between the variables in the model.

**Acknowledgement**
This publication and research has been partially granted by INDESS (Research Universitary Institute for Sustainable Social Development), Universidad de Cádiz, Spain.

***APPENDIX: Questions used in the survey***
**Service quality (SERVIQ)**
SERVIQ1. The formation/training that I received was good.
SERVIQ2. In general, I received sufficient training to be able to work with the system.
SERVIQ3. In general, I´m being supported to be able to work properly with the system.
SERVIQ4. I receive sufficient support to work with the system.
Source: Poelmans et al. (2013).

**General system attributes: reliability (RELIAB)**
RELIAB1. The system is available when I require it.
RELIAB2. The information that I use remains in the system (it doesn´t get lost).
RELIAB3. The system works correctly (it doesn´t get stuck).
Source: Poelmans et al. (2013).

**General system attributes: responsiveness (RESPON)**
RESPON1. The reaction time of the system is correct.
RESPON2. The speed of the system is sufficient for my purposes.
Source: Poelmans et al. (2013).

**General system attributes: integration (INTEGRA)**
INTEGRA1. I can use the system in combination with other tools (word, excel, email…)
INTEGRA2. Tools, such as word, excel, email..., are well integrated into the system
Source: Poelmans et al. (2013).

**BPMS-specific system attributes (SPECIFIC)**
SPECIFIC1. The system allows selecting files/work ítems from the activity received to be able to perform it.
SPECIFIC2. The system (re-)distribute files/work items among your colleagues with the same role.
SPECIFIC3. The system forwards work items to the next step/activity.
SPECIFIC4. The system can put work items back into previous steps .
Source: Poelmans et al. (2013)

**Input quality (INPUTQ)**
INPUTQ1. I have sufficient data entry facilities in the system.
INPUTQ2. I can insert the data in a clear and understandable way (with convenient windows, menu's, fields…).
INPUTQ3. I have sufficient means to correct and/or change the data in the system.
INPUTQ4. I have sufficient help/support when inserting data (e.g. drop down lists, search facilities, pre-entered data…).
INPUTQ5. I can enter data when you need to enter data in the system.
INPUTQ6. I can enter the data in sufficiently detailed way.
Source: Poelmans et al. (2013)

**System quality (SYSTEMQ)**
SYSTEMQ1. The system was easy to learn.
SYSTEMQ2. The system is easy to use.
SYSTEMQ3. The system does what I want it to do (without too much effort).
SYSTEMQ4. The system is user friendly.
SYSTEMQ5. The system is stable (It is tested and does not produce errors).
Source: Poelmans et al. (2013), Wu & Wang (2006) and Rai, Lang, & Welker (2002).

**Information and knowledge quality: retrieval quality (IKQRETRI)**
IKQRETRI1. The information is reliable and accurate.
IKQRETRI2. The information is complete.
IKQRETRI3. The information is readable and easy to understand on the screen.
IKQRETRI4. Electronic presentation/format of the information (on the screen) is adequate.
IKQRETRI5. Printed version/presentation of the information is adequate.
IKQRETRI6. The speed with which the information can be gathered/retrieved is adequate.
IKQRETRI7. The information is updated in the system.
IKQRETRI8. The available information in the system is sufficient for my tasks.
IKQRETRI9. I have sufficient access to the information available in the system.
Source: Poelmans et al. (2013) and Wu & Wang (2006).

**Information and knowledge quality: content quality (IKQCNTN)**
IKQCNTN1. The system makes it easy for me to create knowledge documents.
IKQCNTN2. The words and phrases in contents provided by the system are consistent.
IKQCNTN3. The content representation provided by the system is logical and fit.
IKQCNTN4. The knowledge or information provided by the system is available at a time suitable for its use.
IKQCNTN5. The knowledge or information provided by the system is important and helpful for my work.
IKQCNTN6. The knowledge or information provided by the system is meaningful, understandable, and practicable.
IKQCNTN7. The knowledge classification or index in the system is clear and unambiguous.
Source: Poelmans et al. (2013) and Wu & Wang (2006).

**Information and knowledge quality: context and linkage quality (IKQCNTX)**
IKQCNTX1. The system provides contextual knowledge or information so that I can truly understand what is being accessed and easily apply it to work.
IKQCNTX2. The system provides complete knowledge portal so that I can link to knowledge or information sources for more detail inquire.
IKQCNTX3. The system provides accurate expert directory (link, yellow pages…).
IKQCNTX4. The system provides helpful expert directory (link, yellow pages) for my work.
Source: Poelmans et al. (2013) and Wu & Wang (2006).

**Perceived usefulness for processes (USEFULP)**
USEFULP1. The system is very well suited to do the tasks that it is supposed to do.
USEFULP2. Using the system enables me to handle my cases/work items well.
USEFULP3. In using the system, I can do my tasks in the process more efficiently.
USEFULP4. The system really has added value in the business Process.
Source: Davis (1989) and Poelmans et al. (2013).

**Perceived usefulness for KM (USEFULK)**
USEFULK1. The system helps me acquire new knowledge and innovative ideas.
USEFULK2. The system helps me effectively manage and store knowledge that I need.
USEFULK3. My performance on the job is enhanced by the system.
USEFULK4. The system improves the quality of my work life.
Source: Wu & Wang (2006).

**User satisfaction (SATISF)**
SATISF1. Currently, I am really satisfied with the the system.
SATISF2. I am satisfied that the system meet my knowledge or information processing needs.
SATISF3. I am satisfied with the system effectiveness.
SATISF4. I am satisfied with the system efficiency.
SATISF5. Overall, I am satisfied with the system.
Source: Poelmans et al. (2013); Seddon & Kiew (1996) and Wu & Wang (2006).

**System use for KM (USE)**
USE1. I use the system to help me make decisions.
USE2. I use the system to help me record my knowledge.
USE3. I use the system to communicate knowledge and information with colleague.
USE4. I use the system to share my general knowledge.
USE5. I use the system to share my specific knowledge.
Source: Doll & Torkzadeh (1988) and Wu & Wang (2006).